\documentstyle[twoside,fleqn,epsfig,espcrc2]{article}

\newcommand{\AmS}{{\protect\the\textfont2
  A\kern-.1667em\lower.5ex\hbox{M}\kern-.125emS}}

\title{Fractal Electromagnetic Showers}

\author{L. A. Anchordoqui\address{Department of Physics, 
        Northeastern University, Boston, 
        MA 02115, USA}\thanks{doqui@hepmail.physics.neu.edu}, 
        M. Kirasirova$^a$\thanks{kirasirova@hepmail.physics.neu.edu},
        T. P. McCauley$^a$\thanks{mccauley@hepmail.physics.neu.edu},   
        T. Paul$^a$\thanks{tom.paul@hepmail.physics.neu.edu},
        S. Reucroft$^a$\thanks{stephen.reucroft@cern.ch}, $\,$ and  
        J. D. Swain$^a$\thanks{john.swain@cern.ch}}

\date{}

\begin{document}
\begin{abstract}
We study the self-similar structure of electromagnetic showers
and introduce the notion of the fractal dimension of a shower.
Studies underway of showers in various materials and at various energies
are presented, and the range over which the fractal scaling behaviour
is observed is discussed. Applications to fast shower simulations
and identification, particularly in the context of extensive air
showers, are also discussed. 
\end{abstract}
\maketitle
\section{Introduction}

One of the most serious problems in the analysis of cosmic ray
data is the complex and time-consuming nature of the codes used for 
shower simulation. In order to try to capture
the detailed physics of the processes involved, it is customary
to directly simulate \cite{AIRES,CORSIKA}
the multiplicative branching process whereby
an initial particle gives rise to two or more secondary particles,
each of which, in turn, initiates what is essentially its own
shower, albeit now at lower energy.

Such a process can give rise to large fluctuations, and 
the final distributions of ground particles and their energies
(as well as the longitudinal distribution of the shower as a whole) are
difficult to model with simple parametrizations unless one
is happy to settle for a description of the {\it mean} behaviour
of the shower and forego knowledge of the fluctuations. Indeed,
this is the leading reason that so much Monte Carlo time must be
used for shower simulations: there are no simple analytical 
forms for the relevant distributions which can describe the
fluctuations. The issue is a pressing one for
experiments collecting
large amounts of data which may be difficult to compare against
theory in any form other than a large number of simulated events.

Here we report on the observation that electromagnetic showers
display self-similar behaviour which can be described by a 
multifractal geometry and describe first steps towards formalizing
this concept. Our eventual goal is to describe showers in terms
of what we argue here is the relevant geometry: not one of
smooth functions, but one which allows for irregular geometries
which are better described in terms of fractals. We consider here only
electromagnetic showers, but plan to study hadronic showers
in future work.

\section{Self-Similarity in Electromagnetic Showers}

The idea that an electromagnetic shower should, in some sense,
be a fractal is almost obvious. It is generated recursively from
the two processes"

\begin{enumerate}
\item pair creation: $\gamma\rightarrow {\mathrm{e^+e^-}}$ in the electric
field of a nucleus and;
\item Bremsstrahlung: ${\mathrm{e}}^\pm\rightarrow {\mathrm{e}}^\pm \gamma$
as an electron or positron is deflected by the electric field of a nucleus
\end{enumerate}

This is illustrated in figure 1 which shows the particles
making up a shower produced
by a 100 GeV electron entering a block of aluminum 150 cm long 
(radiation length 8.9 cm) as simulated using the {\sc geant4} 
program \cite{GEANT4}.

Each final state particle from an interaction effectively initiates its own
electromagnetic shower, and each process has a similar cross section
to occur in matter. As long as the energies involved are large
compared with the energy required to create an electron-positron 
pair (and thus also large compared to atomic processes such as
ionization), each step is much the same as the one  before it, 
but at a reduced energy.

Figure 2 shows a slice through the block right at the
far end with the
point of intersection of each particle with the slice shown
as a black dot whose radius is independent of energy. Here one clearly
sees the shower core, with a diminishing density of particles
with distance from the centre.

\begin{figure}[htbp]
\begin{center}
\hspace*{-5mm}\mbox{\epsfig{file=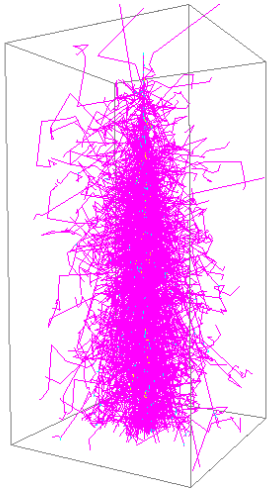,width=.275\textwidth,clip=}}
\caption{Three-dimensional view of the tracks making up an
electromagnetic shower due to a 100 GeV electron entering an
aluminum block 150 cm in length.}
\end{center}
\label{fig:shower}
\end{figure}
\begin{figure}[htbp]
\begin{center}
\hspace*{-5mm}\mbox{\epsfig{file=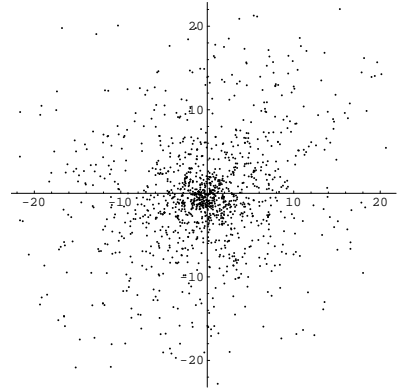,width=.32\textwidth,clip=}}
\caption{Points of intersection of particles with a slice 150 cm 
from the entry point of a 100 GeV electron into an aluminum block.
Axes are perpendicular to the shower axis and are marked in centimetres.}
\end{center}
\label{fig:slice}
\end{figure}
\vspace*{-15mm}
\section{Fractals and Multifractals}

There are many ways to characterize self-similar objects, but
the most common and well-known way is in terms of fractal dimensions. There are
many different concepts of fractal dimension which are useful, and perhaps
the most obvious is that of mass dimension, $D_M$. The idea here
is to see how the total energy $E_{TOT}(R)$ (considered now as a sort of 
weight)
within a disk of radius $R$ varies as $R$ is changed. If the
distribution were one-dimensional (a line of uniform energy deposited in the 
plane),
one would find 

\begin{equation}
E_{TOT}(R) \propto R^{1} 
\end{equation}

\noindent and one would take the exponent in the foregoing equation to be
the dimension of the distribution.

If the energy were uniformly distributed over the whole plane, one would
find 

\begin{equation}
E_{TOT}(R) \propto R^{2} 
\end{equation}

\noindent and conclude again that the exponent in the scaling law for the
energy should be interpreted as the dimension of the distribution.

In the event that a scaling law of the form $E_{TOT}(R)\propto R^{D_M}$ holds
for a non-integer $D_M$, we call $D_M$ the ``fractal mass dimension''.
A plot of $\log(E)$ as a function of $\log(R)$ will then have a slope
in the limit of small $R$ which is $D_M$.

Two points are important to keep in mind here: first that there are
no true fractals in nature as there are always some smallest and largest
value for variables in the problem beyond which scaling behaviour does not
hold, and second that one must be careful to watch for systematic effects
which can bias estimates of the dimension. Systematic effects which 
we have had to be wary of include the fact that early in the shower
development
the central core can contain particles which carry a large fraction of the
initial energy and give the radial energy distribution a spike at small
$R$ which does not correspond to scaling behaviour.

\begin{figure}[htbp]
\begin{center}
\hspace*{-5mm}\mbox{\epsfig{file=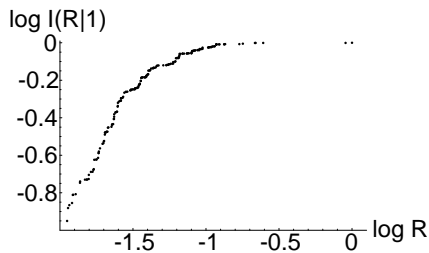,width=.38\textwidth}}
\caption{Summed energy as a function of radius from the
shower centre. Logarithms are base 10.}
\end{center}
\label{fig:P1}
\end{figure}

In the case of the electromagnetic shower with the slice taken
at the end of the shower at 150 cm, we look at the summed energy
(scaled so that the total energy is 1)
as a function of the fraction of the radius out (scaled so that the
maximum radius is 1). This quantity we denote as $I(R|1)$ for reasons
which will become clear later in the text.
Plotting logarithms against logarithms
(base 10), we find the distribution shown in figure \ref{fig:P1}.
The first thing to notice is that the curve is reasonably
approximated by a straight line at small radii.
The second thing
to notice is that the whole curve is {\it not} a 
straight line. At large radii we start to reach the physical 
boundaries of the shower and cannot expect scaling to hold.

In fact, even at very small radii, there is some anomalous
structure which can be traced to the effects of very energetic
particles very close to the core, which give an additional spike
of energy to the distribution which cannot be expected
to be a part of any overall scaling behaviour. This effect is
more pronounced earlier in the shower.

The scaling properties
of the shower are thus different in different parts of the plane, and
in order to quantify this further, we study the scaling behaviour of
cumulative moments of the energy distribution defined for $q>0$ by

\begin{equation}
I(R|q) = \frac{\sum_{r<R} E_i^q}{\sum_{\mathrm{all\ }i}E_i^q} 
\end{equation}
\noindent where $E_i$ are the energies contained in a disk going out to radius $R$
and the sum is taken over all particles
within a distance $r<R$. What units are used is not important as we are only interested
in the average scaling behaviour of the curves at small $R\rightarrow 0$. 
(As discussed
earlier in the text, the region of very small $R$ should be avoided for physical reasons,
and we will avoid the subtleties of precise numerical analyses in this short communication.) 
For graphical
purposes here, $R$ is normalized so that the particle with the largest radial
distance out is at $R=1$ and the moments are defined so that their value
at maximum radius is unity. We can then introduce an infinite family\cite{fractalrefs}
of fractal dimensions $D_q$ defined for $q>0$ by

\begin{equation}
D_q = \lim_{R\rightarrow 0} \left< \frac{1}{q}\frac{\partial \log I(R|q) }{\partial \log R} \right>
\end{equation}

\noindent with the understanding that the limit must still lie in the scaling region in 
physical examples.

Figure 4 shows the scaling behaviour of moments of the
electromagnetic shower corresponding to how the sums of the squares
and cubes of the energy grow with distance. 
\begin{figure}[htbp]
\begin{center}
\hspace*{-5mm}\mbox{\epsfig{file=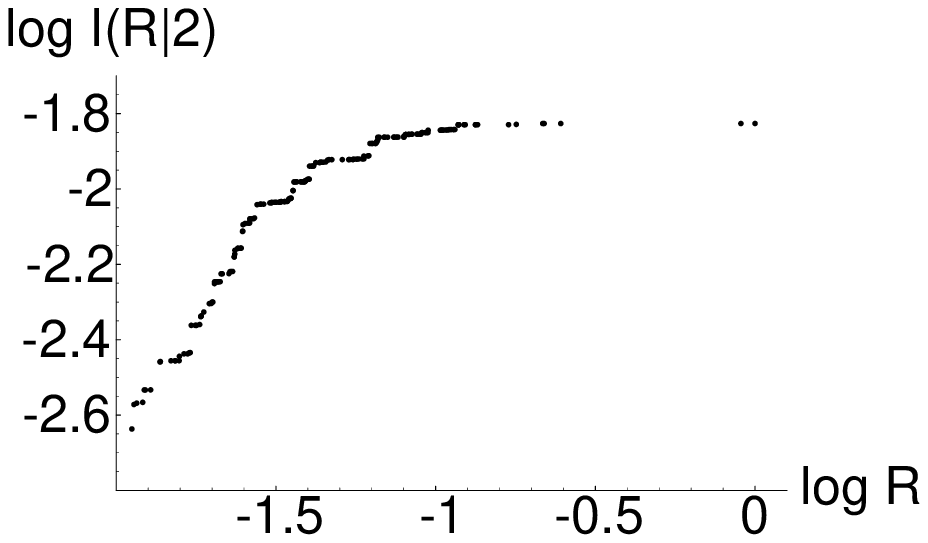,width=.38\textwidth,clip=}} \\
\mbox{\epsfig{file=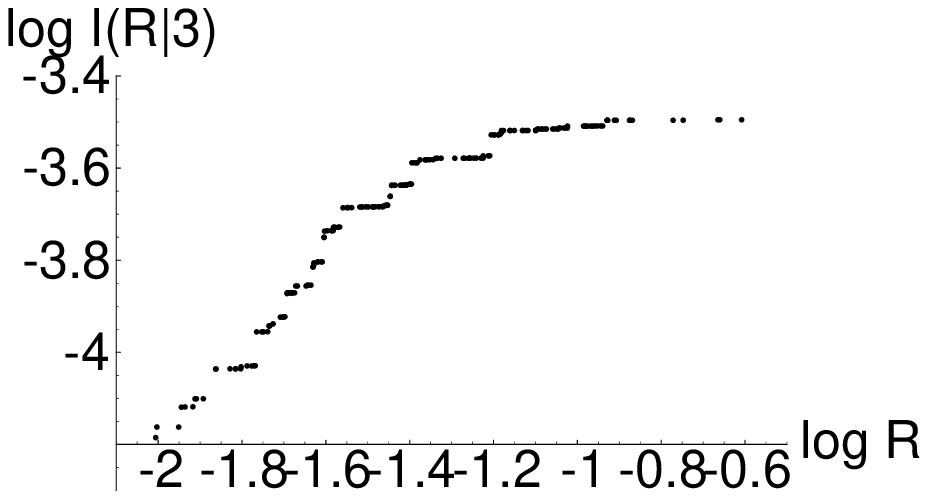,width=.4\textwidth,clip=}}
\caption{Total energy-squared (above) and total energy-cubed 
as functions of radius from the shower centre (see text for normalizations).
Logarithms are base 10.}
\end{center}
\label{fig:Ptwo}
\end{figure}
For a homogeneous and uniform fractal structure we expect the $D_q$ to be
equal. If not, then
we describe the distribution as multifractal in that it requires more than
one fractal dimension in order to characterize it.  
The associated $D_q$ for small $q$ estimated from finite differences
in the scaling region are all approximately equal within the
errors in the data here and approximately unity, suggesting a good
degree of homogeneity. 
It is important to keep in mind that
the results in this paper are presented for a full, realistic GEANT
simulation, and include ionization, delta-ray, and other soft processes,
so some care is needed in interpreting the results as if they corresponded
to a pure electromagnetic shower generated only by pair creation and
Bremsstrahlung (which is, of course, not realizable
in nature).

The definition of fractal
dimensions can also be continued to $q\leq 0$, but this has some subtleties involved
with the fact that as $q\rightarrow\infty$
the highest energy particles contribute most, while as $q\rightarrow -\infty$
the lower energy ones dominate. 
In particular, some care must be used with the $D_q$ for
$q<0$ as they give high weights to softer particles which are not part of the hard
shower process. These matters,
as well as more precise results on dimensions including energy and
material dependence will be presented elsewhere\cite{inprep}.

\vspace*{-2mm}
\section{Further Work}

Clearly space limitations make it impossible to cover the material
as completely as one would like, but several points concerning
work not discussed here are worth making.
First of all, we expect fractal behaviour in all three dimensions,
and in this discussion we have neglected the longitudinal scaling
behaviour, where the full shower is made of many scaled and translated
showers superimposed along the shower axis. In addition, there are clearly
angular correlations and fluctuations, and studies can be made at a given
fixed radius of
the scaling behaviour of the shower as a function of the angular coordinate
which we have integrated out in this discussion. The relation of these
ideas to the concept of intermittency, especially as studied in hadronic 
jets has not escaped our notice and is currently under investigation.

One of the main goals of this work is to better understand the
geometry of electromagnetic (and other) showers in order to try to
parametrize them by the appropriate non-smooth basis functions, such 
as wavelets. Such a parametrization should allow the fast generation
of showers without the attendant loss of information concerning
large fluctuations\cite{inprep} which has so far been handled only by the use of
enormous computational resources.

\vspace*{-2mm}
\section{Acknowledgements}

We would like to thank the US National Science Foundation 
and CONICET, Argentina for support. We would also like
to thank our collaborators on the Pierre Auger Project,
as well as on L3 and CMS for useful discussions on 
electromagnetic calorimetry.

\end{document}